\documentclass[reprint,prl,superscriptaddress]{revtex4-2}
\usepackage{graphicx}

\begin{document}

\title{High resolution transcranial imaging based on the optoacoustic memory effect}

\author{X. Lu\'is De\'an-Ben} \email{Corresponding author: xl.deanben@pharma.uzh.ch\\} 
\affiliation{Institute for Biomedical Engineering and Institute of Pharmacology and Toxicology, Faculty of Medicine, University of Z\"urich, Z\"urich, Switzerland.}
\affiliation{Institute for Biomedical Engineering, Department of Information Technology and Electrical Engineering, ETH Z\"urich, Z\"urich, Switzerland.}
\author{Daniel Razansky}
\affiliation{Institute for Biomedical Engineering and Institute of Pharmacology and Toxicology, Faculty of Medicine, University of Z\"urich, Z\"urich, Switzerland.}
\affiliation{Institute for Biomedical Engineering, Department of Information Technology and Electrical Engineering, ETH Z\"urich, Z\"urich, Switzerland.}

\date{\today}

\begin{abstract}

Acoustic impedance mismatches between soft tissues and bones are known to result in strong aberrations in optoacoustic and ultrasound images. Of particular importance are the severe distortions introduced by the human skull, impeding transcranial brain imaging with these modalities. While modelling of ultrasound propagation through the skull may in principle help correcting for some of the skull-induced aberrations,  these approaches are commonly challenged by the highly heterogeneous and dispersive acoustic properties of the skull and lack of exact knowledge on its geometry and internal structure. Here we demonstrate that the spatio-temporal properties of the acoustic distortions induced by the skull are preserved for signal sources generated at neighboring intracranial locations by means of optoacoustic excitation. This optoacoustic memory effect is exploited for building a three-dimensional model accurately describing the generation, propagation and detection of time-resolved broadband optoacoustic waveforms traversing the skull. The memory-based  model-based inversion is then shown to accurately recover the optical absorption distribution inside the skull with spatial resolution and image quality comparable to those attained in skull-free medium. 

\end{abstract}

\maketitle 

Imaging modalities generally rely on the information carried by a certain type of wave propagating through the imaged object. Ultrasound (US) waves traveling through biological tissues have been extensively used in biomedical applications to visualize internal structures of living organisms in a fully non-invasive manner \cite{rix2018advanced,deffieux2018functional}. Scattering, attenuation, speed of sound variations and other elastic properties can be mapped with US provided wave propagation is well known or can be properly modelled \cite{lafci2020noninvasive}. In the last two decades, optoacoustic (OA, photoacoustic) imaging has further emerged as a new modality capitalizing on the weak scattering of US in biological tissues (compared to photons) to map the optical absorption contrast \emph{in vivo}, thus providing otherwise unattainable functional and molecular information \cite{dean2017advanced,yao2018recent}. OA image reconstruction is commonly performed with analytical or iterative algorithms derived by assuming uniform acoustic properties and undisturbed US propagation \cite{rosenthal2013acoustic}. More sophisticated algorithms accounting for acoustic aberrations have further been developed \cite{dean2013weighted,cox2017modeling,haltmeier2019reconstruction}. However, much like for pulse-echo US, the presence of bones, lungs or other tissues whose acoustic properties strongly differ from soft tissues massively challenges image formation.

In the last decade, newly-developed wavefront shaping methods have demonstrated that the complex optical wavefronts resulting from strong light scattering can actually be shaped and focused \cite{horstmeyer2015guidestar,vellekoop2015feedback}. Indeed, the apparently random speckle light patterns observed when coherent light propagates through a scattering sample encode valuable information on the original light source(s). The scattering process is too complex to be theoretically modelled but can be described instead as a transmission matrix experimentally built from a set of known incident wavefronts \cite{chaigne2014controlling}. Knowledge of this matrix enables full control of the light beam, e.g. facilitating high resolution imaging through a scattering sample. However, building the transmission matrix is a slow process and not always feasible. A key phenomenon affecting the wavefront transmitted through a scattering layer is the so-called memory effect. This refers to the fact that the speckle patterns corresponding to plane wavefronts within a range of impinging angles are highly correlated and spatially shifted \cite{feng1988correlations,freund1988memory}. The memory effect implies that all wavefronts undergo the same distortion, while the spatial shift is associated with the phase distribution corresponding to the propagation direction.

The memory effect is not exclusive to light as the propagation of other types of waves through scattering media can be similarly described \cite{mosk2012controlling}. In biomedical US, the human skull arguably represents the most acoustically distorting layer in the human body. Transcranial US propagation involves not only scattering events but also reverberations and mode conversions of the incident longitudinal wave to shear or guided waves \cite{estrada2018observation}. We hypothesized that the acoustic aberrations induced by the skull are retained by signal sources at neighboring locations. Based on this memory effect, we define an OA transmission model to be used to recover the absorption distribution from the collected time-resolved transmitted signals.

OA signal excitation is commonly done with short laser pulses fulfilling the so-called stress and thermal confinement regimes. Absorption of the laser energy causes an initial pressure rise followed by propagation of an US (pressure) wave. Assuming a negligible fractional volume change and a sufficiently short (Dirac delta) $\delta(t)$ pulse, the generation and propagation of the pressure wave $p(r,t)$ is governed by \cite{cox2007k}

\begin{equation} \label{Eq1}
	\frac{\partial^2p(r,t)}{\partial{}t^2} - c(r)^2\rho(r)\nabla\cdot\left(\frac{1}{\rho(r)}\nabla{}p(r,t)\right) = \Gamma(r)H(r)\frac{\partial\delta(t)}{\partial{}t},
\end{equation}

where $c(r)$ and $\rho(r)$ are the speed of sound and mass density of the medium, $\Gamma(r)$ is the dimensionless Grueneisen parameter and $H(r)$ is the thermal energy deposited per unit volume. The term $\Gamma(r)H(r)$ corresponds to the initial pressure rise $p_{0}(r)$. Note that Eq.~\ref{Eq1} assumes a non-attenuating heterogeneous medium only sustaining longitudinal waves. Modeling of bones and other hard tissues implies an additional consideration of elastic wave propagation in solids, US absorption terms and appropriate boundary conditions at the interfaces between the propagation domains. It is important to highlight that the well-defined source term in Eq.~\ref{Eq1} establishes the bipolar temporal shape and the broad frequency spectrum of the generated US wave. This represents a fundamental difference of OA with respect to pulse-echo US since a general mathematical description of back-scattering sources is not possible. Eq.~\ref{Eq1} thus serves as a basis for deriving OA reconstruction algorithms. For example, if the OA sources are confined within a defined region enclosed by a Cartesian grid of voxels, Eq.~\ref{Eq1} can be approximated via

\begin{equation} \label{Eq2}
	\frac{\partial^2p(r,t)}{\partial{}t^2} - c(r)^2\rho(r)\nabla\cdot\left(\frac{1}{\rho(r)}\nabla{}p(r,t)\right) \approx \frac{\partial\delta(t)}{\partial{}t}\sum_{i=1}^{N}p_{0,i}k_{i}(r),
\end{equation}

where $p_{0,i}$ is the initial pressure at the center of the $i$-th voxel of the grid and $k_{i}(r)$ represent the corresponding interpolation kernel. The general solution of Eq.~\ref{Eq2} can be expressed via

\begin{equation} \label{Eq3}
	p(r,t) = \sum_{i=1}^{N}p_{0,i}p_{i}(r,t),
\end{equation}

being $p_{i}(r,t)$ the solution of Eq.~\ref{Eq2} for a unit initial pressure rise at the $i$-th voxel. Considering pressure values at a set of points and instants, expressed as a column vector $\mathbf{p}$, Eq.~\ref{Eq3} can be expressed as a function of the initial pressure rise at the voxels of the image grid, also expressed as a column vector $\mathbf{p}_{0}$, i.e.

\begin{equation} \label{Eq4}
	\mathbf{p} = \mathbf{A}\mathbf{p}_{0}.
\end{equation}

\begin{figure}
	\centering
		\includegraphics[width=0.45\textwidth]{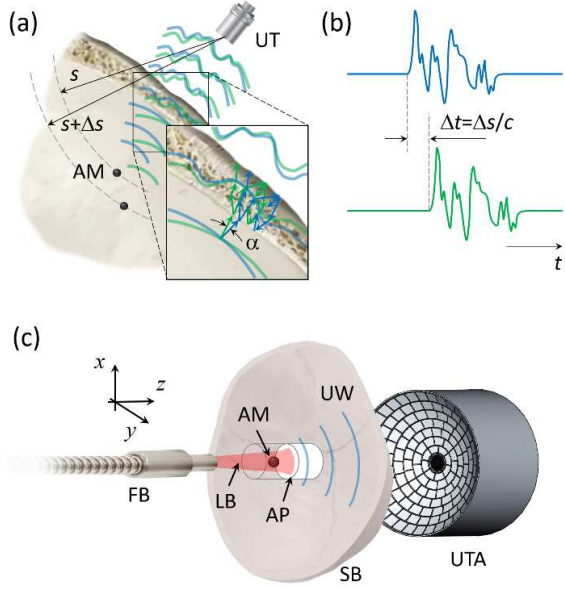}
		\caption{Basic principle and experimental set-up for validating the optoacoustic (OA) memory effect. (a) OA responses generated by closely located absorbing microspheres (AM) are shown in green and blue, respectively. The propagating paths of the waves are approximately the same for a small difference in the angle of incidence $\alpha$ (see inset). (b) Time-resolved OA signals recorded by the ultrasound transducer (UT) for the two indicated positions of the AM. The difference in time-of-flight ($\Delta{}t$) relates to the speed of sound (c) in the medium. (c) Lay-out of the experimental system employed to verify the existence of the OA memory effect for a human skull specimen. FB – fiber bundle, AM – absorbing microsphere, LB – laser beam, AP – agar phantom, UW – ultrasound wave, SB – skull bone, UTA – ultrasound transducer array.}
	\label{Fig1}
\end{figure}

Each column of the model-matrix $\mathbf{A}$ corresponds to the pressure values for a unit initial pressure rise at the corresponding voxel on the grid. These can be measured experimentally e.g. by scanning an OA source across the field of view \cite{dean2019acoustic,li2020snapshot}, a very slow and not always feasible process. On the other hand, $\mathbf{A}$ can be built theoretically if an analytical solution of Eq.~\ref{Eq2} is available, for instance in the case of a uniform acoustic medium \cite{rosenthal2013acoustic}. This can be achieved by taking as a reference a column of the model matrix, i.e., the bipolar signal corresponding to the interpolation kernel of a given voxel. The rest of the signals (columns of $\mathbf{A}$) can be obtained by appropriately delaying and scaling the reference signal \cite{ding2020model}. This implies the existence of an intrinsic OA memory effect maintaining the temporal shape of OA waves propagating through a uniform acoustic medium. However, OA waves traveling through a highly acoustically mismatched medium like the human skull experience severe acoustic aberrations (Fig.~\ref{Fig1}a) manifested as complex time profiles of the collected signals (Fig.~\ref{Fig1}b). Longitudinal waves impinging upon the skull generally undergo refraction as well as mode conversion to shear and guided waves, followed by attenuation, reflections of subsequent mode conversions in the opposite interface and internal layers (inset in Fig.~\ref{Fig1}a). The overall distortion of the waves depends on the propagating path through the skull, which in turn depends on the shape of the incident beam and the angle of incidence. OA waves generated at various locations generally propagate along different paths. Yet, the shape and angle of incidence of waves generated at neighboring positions, which are also sufficiently distant from the skull interface, are expected to very similar. Thereby, a high level of similarity is expected in their propagating paths and the associated distortions. This would result in the time delay corresponding to a difference in time-of-flight being the only major difference between the transmission of OA signals generated at relatively close locations (Fig.~\ref{Fig1}b). 

We experimentally verified the existence of the OA memory effect by measuring OA signals transmitted through a human skull. The lay-out of the experimental system is depicted in Fig.~\ref{Fig1}c. Basically, an output beam of optical parametric oscillator (OPO)  laser (Innolas GmbH, Krailling, Germany) tuned to $720\;\mathrm{nm}$ wavelength and operated at $50\;\mathrm{Hz}$ pulse repetition frequency (PRF) was coupled to a bundle consisting of $480$ individual fibers (CeramOptec GmbH, Bonn, Germany). The beam was directed onto a cylindrical agar phantom embedding an absorbing $\sim{}100\;\mu{}\mathrm{m}$ diameter micro-sphere (Cospheric BKPMS $90–106$, Santa Barbara, USA). The phantom further contained a highly concentrated layer of Intralipid solution causing scattering and thus attenuation of the light beam. The generated OA signals were collected with a spherical US matrix array (Imasonic SaS, Voray, France) having $40\mathrm{mm}$ radius, $90^\circ$ angular coverage and comprising of $256$ individual piezocomposite elements with $4\mathrm{MHz}$ central frequency and $\sim{}100\%$ $-6\;\mathrm{dB}$ bandwidth \cite{dean2013portable}. The microsphere was first  positioned at the center of the array and then scanned along the three Cartesian axes in a stepwise manner with the OA signals acquired at each position (averaged for $100$ consecutive laser pulses). Subsequently, an excised human skull was placed between the phantom and the array (Fig.~\ref{Fig1}c). The distance from the sphere to the skull was $\sim{}1\;\mathrm{cm}$ and the thickness of the skull at the closest point was $\sim{}1.5\;\mathrm{mm}$. A fresh-frozen fronto-temporal bone samples derived from decompressive hemicraniectomy was collected according to protocols established by the ethics committee of the Department of Neurosurgery at the University Hospital Cologne. The skull sample was kept at $–80^\circ{}C$ and was later degassed for $3$ hours and immersed in $0.9\%$ saline solution during the imaging experiments. The phantom was then re-scanned with the OA signals acquired and averaged at the same scanning positions.

\begin{figure}
	\centering
		\includegraphics[width=0.45\textwidth]{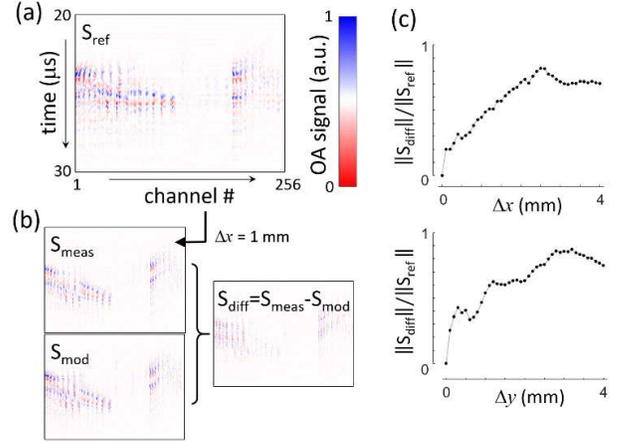}
		\caption{Experimental validation of the OA memory effect. (a) Measured sinogram ($\mathrm{S}_{\mathrm{ref}}$) showing the time-resolved OA waveforms recorded by all the $512$ channels of the spherical matrix array transducer in response to the laser pulse absorption by an absorbing microparticle located at the center of the array. (b) Measured sinogram for a $1\;\mathrm{mm}$ lateral shift of the particle ($\mathrm{S}_{\mathrm{meas}}$) along with the respective sinogram calculated with the memory-effect-based OA model ($\mathrm{S}_{\mathrm{mod}}$). The difference between the two sinograms is also shown. (b) Relative norm of the difference between measured and calculated sinograms as a function of the lateral shift of the absorbing microparticle.}
	\label{Fig2}
\end{figure}

In the sinogram showing the time-resolved OA signals for the micro-sphere positioned at the center of the array (Fig.~\ref{Fig2}a) most signals are severely distorted from their original bipolar shape. It is also observed that some channels exhibit very low signal amplitude, arguably corresponding to reflections and/or mode conversion to guided waves. When the microsphere is displaced by $1\;\mathrm{mm}$ in the lateral direction, the measured sinogram can be reasonably predicted by considering the OA memory effect (Fig.~\ref{Fig2}b). A more quantitative comparison is provided by calculating the relative norm of the difference between the measured and theoretical sinograms as a function of the lateral shift of the microsphere (Fig.~\ref{Fig2}c). It is shown that the sinograms calculated by considering the OA memory effect at least partially correlate with the measured ones for lateral shifts of up to $\sim{}3\;\mathrm{mm}$. As expected, such correlation depends on the scanning direction due to the heterogeneous skull properties.

\begin{figure}
	\centering
		\includegraphics[width=0.45\textwidth]{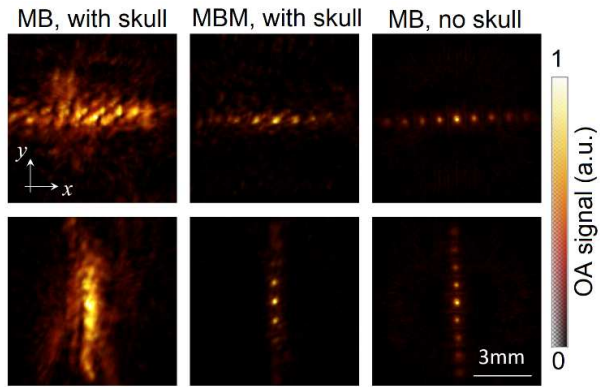}
		\caption{Transcranial OA imaging of absorbing microspheres. (left) Volumetric OA reconstructions of absorbing $100\;\mu\mathrm{m}$ diameter microspheres obtained by means of standard 3D model-based (MB) reconstruction. The images were rendered by superimposing the signals from $9$ individual microspheres separated by $1\;\mathrm{mm}$ along the lateral directions. Maximal intensity projections (MIPs) along the axial ($z$) dimension are shown. Prior to the reconstruction, the raw signals were band-pass filtered between $0.1$ and $8\;\mathrm{MHz}$ to remove low frequency offsets and high-frequency noise. (middle) Corresponding reconstruction obtained with the memory-effect-based model (MBM). (right) Reconstructed MB image in absence of the skull.}
	\label{Fig3}
\end{figure}

We subsequently tested the performance of the memory-effect-based OA inversion approach, for which the columns of the model matrix in Eq.~\ref{Eq4} are calculated by considering the experimental signals for a given microsphere position. Specifically, OA image reconstruction involves minimizing the energy functional defined via

\begin{equation} \label{Eq5}
	\mathbf{p}_{0,\mathrm{sol}} = \mathrm{argmin}_{\mathbf{p}_{0}}\left\{\left\|\mathbf{p}_{m}-\mathbf{A}\mathbf{p}_{0}\right\|_{2}^{2}+\lambda\left\|\mathbf{p_0}\right\|_{2}^{2}\right\},
\end{equation}

where $\mathbf{p}_{m}$ are the measured signals and $\lambda$ is a regularization parameter selected by maximizing curvature of the L-curve \cite{hansen1998rank}. When inspecting an OA image of several microspheres obtained with the standard model-based (MB) approach assuming a homogenous non-attenuating acoustic medium \cite{ding2020model}, the reconstruction is clearly affected by strong artefacts caused by the acoustic distortion of waves propagating through the skull (Fig.~\ref{Fig3}, left). The equivalent image reconstructed with the memory-effect-based model (MBM) (Fig.~\ref{Fig3}, middle) enables clearly distinguishing the individual spheres and visualizing their distribution, matching well the MB-reconstructed image obtained with the skull absent, at least for the microspheres locations within a range of approximately $\pm{}2\;\mathrm{mm}$ from the center (Fig.~\ref{Fig3}, right).

\begin{figure}
	\centering
		\includegraphics[width=0.45\textwidth]{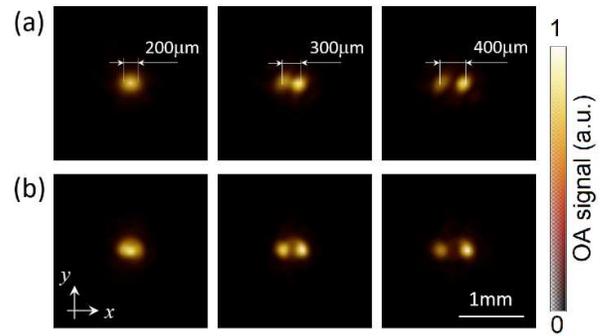}
		\caption{Spatial resolution characterization. (a) Images of two closely located $100\;\mu\mathrm{m}$ diameter microspheres reconstructed with the memory-effect-based OA model through the skull. (b) Equivalent images reconstructed with the standard OA model without the presence of the skull.}
	\label{Fig4}
\end{figure}

We further evaluated the achievable spatial resolution performance limits of the method by reconstructing two microspheres separated by $200$, $300$ and $400\;\mu\mathrm{m}$. The microspheres could be clearly distinguished through the skull when separated by at least $300\;\mu\mathrm{m}$ with the memory-effect-based OA model (Fig.~\ref{Fig4}a). Similar results are obtained when considering the standard model-based inversion without the skull (Fig.~\ref{Fig4}b), in congruence with the expected spatial resolution of the matrix array probe \cite{dean2013portable}.

\begin{figure}
	\centering
		\includegraphics[width=0.45\textwidth]{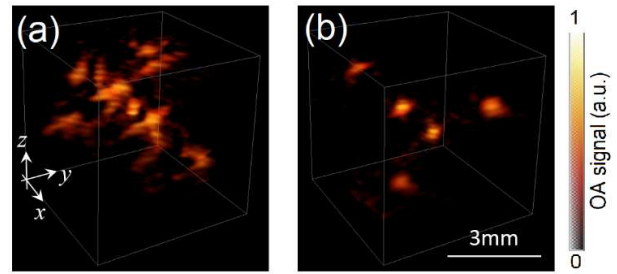}
		\caption{Transcranial imaging of a three-dimensional random distribution of point absorbing targets. (a) Transcranial OA image of five randomly distributed $100\;\mu\mathrm{m}$ diameter microspheres reconstructed with the standard model-based inversion. (b) The corresponding image reconstructed with the memory-effect-based OA model.}
	\label{Fig5}
\end{figure}

Finally, we tested the capability of the memory-effect-based OA model to accurately reconstruct a random distribution of microspheres in a three-dimensional region. As expected, the image reconstructed with the standard OA model clearly manifests the skull-induced distortions (Fig.~\ref{Fig5}a), whilst the microspheres are accurately resolved in the image reconstructed with the OA memory-effect-based model (Fig.~\ref{Fig5}b). 

The distortions introduced by acoustically mismatched tissues represent a major hinderance for OA imaging of certain regions such as brain and lungs. Particularly, major efforts have been dedicated to modeling transcranial US propagation in humans \cite{poudel2020iterative}, which is hampered by the heterogeneous elastic and dimensional properties of the skull and the complex mode conversions between longitudinal, shear and guided waves \cite{estrada2018observation}. We have shown that these effects are locally preserved, which could be exploited for building a linear model enabling accurate OA image reconstruction via iterative inversion. The quality of images obtained through a full thickness human skull was massively enhanced by assuming an inversion model accounting for the OA memory effect in comparison to the standard OA reconstruction algorithm assuming a uniform acoustic medium. This anticipates the general applicability of the suggested approach in other regions of the body where bones, lungs or other tissues with strongly mismatched acoustic properties are present. Transcranial propagation is of particular importance considering the powerful neuroimaging capabilities of OA imaging in small rodents \cite{ovsepian2017pushing,gottschalk2019rapid}. Recently, OA mapping of human brain activity has also been demonstrated in post-hemicraniectomy patients \cite{na2021massively}. The good agreement with the activity readings provided by functional magnetic resonance imaging (fMRI) anticipates that OA can be used to advance our understanding on human brain function in health and disease provided high resolution transcranial imaging is enabled.

The reconstruction approach suggested in this work relies on OA signals generated by a point-like source that serves as the guiding star for the memory-effect-based reconstruction. Creating such point sources may turn challenging in many practical cases, particularly in the clinical setting. Insertion of a thin optical fiber with a light-absorbing particle attached to its tip may potentially be used for this purpose \cite{noimark2018polydimethylsiloxane}, which might be considered a minimally-invasive procedure similar to a needle puncture routinely used in the clinics for biopsies and other interventions. An alternative approach to collect the required transcranial signals from point-like OA sources consists in intravenous injection of nano- or micro-particles having sufficient absorption to be individually detected. The feasibility to detect individual microparticles sufficiently small to flow throughout the vascular network has recently been demonstrated, which has further enabled breaking through the acoustic diffraction barrier via the so-called localization OA tomography (LOT) \cite{dean2020noninvasive}. Another potential method to experimentally measure the OA signals corresponding to a point source within the region of interest is via combination of OA and pulse-echo US. Back-scattered US wavefronts from individual bubbles have successfully been used for correcting for transcranial propagation effects \cite{demene2021transcranial}. Such echoes may be used for estimating the OA signals at the same location, provided the delays and differences in the acoustic spectra are properly accounted for. Alternatively, point US sources outside the sample may serve to generate echoes corresponding to “virtual” sources within the region of interest, which could also potentially be used to estimate the required OA signals.

The use of the OA memory effect comes with certain limitations. On the one hand, propagation through acoustically mismatched regions does not only imply wavefront distortion but also acoustic attenuation reducing the signal to noise ratio (SNR) of the collected signals. For example, transcranial US propagation in human adults typically results in a signal loss of $80-90\%$ \cite{na2021massively}. On the other hand, the memory effect remains valid only when the waves propagate along very similar paths. For coherent plane waves, the range of validity of the memory effect is known to depend on the thickness of the scattering layer relative to the wavelength \cite{freund1988memory}. Location of the imaged region of interest with respect to the skull interface is also expected to have an influence. These limitations establish trade-offs between the optimal bandwidth of the US transducer(s) employed, achievable spatial resolution and field of view.

In conclusion, the OA memory effect is expected to facilitate imaging through biological tissues with strongly mismatched acoustic properties. Transcranial imaging based on this effect may thus enable the clinical translation of OA modality toward brain imaging applications.

\textbf{Acknowledgments} This study was partially supported by the European Research Council Consolidator grant ERC-2015-CoG-682379. We would like to thank Dr. Volker Neuschmelting for providing the skull samples.


\end{document}